# Time transfer by laser link between China and France

C. Zhao [1, 2], W.-T. Ni [1], E. Samain [3]

[1]Center for Gravitation and Cosmology, Purple Mountain Observatory, Chinese Academy of Sciences, Nanjing, 210008 China.

Email: zhaocheng@pmo.ac.cn, wtni@pmo.ac.cn.

[2]Graduate University of Chinese Academy of Sciences, Beijing, 100049 China.

[3]Observatoire de la Côte d'Azur, UMR Gemini, 06460 Caussols, France.

Email:etienne.samain@obs-azur.fr

**Abstract.** To advance from milli-arcsecond to micro-arcsecond astrometry, time keeping capability and its comparison among different stations need to be improved and enhanced. The T2L2 (Time transfer by laser link) experiment under development at OCA and CNES to be launched in 2008 on Jason-2, allows the synchronization of remote clocks on Earth. It is based on the propagation of light pulses in space which is better controlled than the radio waves propagation. In this paper, characteristics are presented for both common view and non-common view T2L2 comparisons of clocks between China and France.

1. Introduction

T2L2 on Jason-2 will permit to synchronize remote ground clocks and compare their frequency stabilities using laser telemetry with a performance 1-2 orders of magnitude better than present. T2L2 will allow to perform a synchronization of a ground clock and a space clock, and to measure the stability of remote ground clocks over continental distances with 1 ps over 1000 s. China possesses 5 fixed laser ranging stations and a mobile station -- TROS; their geography allows common view comparisons among each other, and non-common view comparisons between a Chinese station and a French station.

2. T2L2 principle and performance budgets

A given light pulse is emitted from station A at time $t_s$, and received in Jason-2 at time $t_b$; this pulse is also bounced from the retroreflector on Jason-2 and received at station A at time $t_r$. The synchronization $X_{AS}$ between the ground clock A and the satellite clock S is:

$$X_{AS} = (t_s + t_r)/2 - t_b + \tau_{Geometry} + \tau_{Atmosphere} + \tau_{Relativity}$$

Also we can get $X_{BS}$, the synchronization between another ground clock B and the satellite clock. So, the time transfer between A and B can simply defined by:

$$X_{AB} = X_{AS} - X_{BS}$$

The performance budget is show in fig 1. In common view configuration, T2L2 should reach the majority of current atomic clocks (including cold atoms) for integration times exceeding 1000 s. In non-common view, with the limitation imposed by the onboard clock T2L2 will still offer an interesting alternative in calibration campaigns of radiofrequency frequency and time transfer systems based on transportable stations.

.

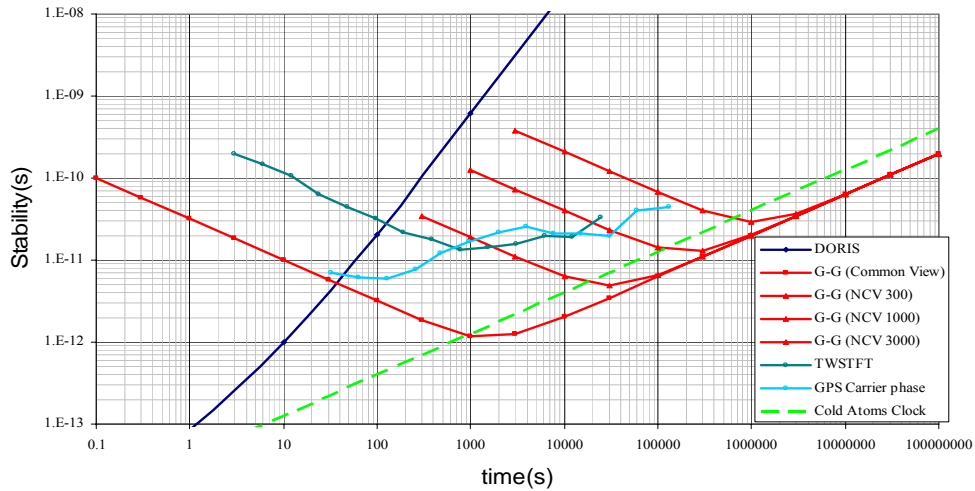

**Figure 1**. T2L2 ground-to-ground time stability in common view (CV) and non-common view (NCV) configuration with various interval ("NCV 300" means non-common view comparison with 300 s interval between two stations).

## 3. Time transfer between Grasse and Chinese stations

T2L2 is a passenger instrument of Jason-2. Jason-2 will be launched by Delta II 7320 into an earth orbit with period 6745.72 s, inclination 66° and apogee 1,336 km high. The orbit drifts to the east 39.5° every revolution and returns to the original orbit every cycle (9.9156 days). This orbit will allow common view comparisons with about 3000 km baseline, with 6 repeating ground tracks per day over the ground stations.

For common view time transfer, all five fixed laser ranging stations in China are able to track the Jason-2 synchronously with a long common duration. We could expect abundant time transfer results in common view between these stations.

For non-common view time transfer, as Table 1 shows, there are around 20 good passes that each Chinese station can access Jason-2 after several minutes when Grasse station loses its sight. This is quite good for non-common view time transfer.

Table.1 Grasee (France) – Chinese Stations Time Transfer in Non Common View. (Interval means time separation between the ending of one station and the start of that of the other one.)

| Station | Total Access | Good Access | Average duration (sec) Grasse | Average duration (sec) China | Interval (sec) Average | Interval (sec) Maximum | Interval (sec) Minimum |
|---|---|---|---|---|---|---|---|
| Grasse-Beijing | 35 | 25 | 972.92 | 923.11 | 392.94 | 554.95 | 345.52 |
| Grasse-Changchun | 32 | 24 | 986.73 | 973.47 | 401.20 | 620.41 | 345.81 |
| Grasse-Kunming | 26 | 18 | 958.15 | 897.78 | 536.88 | 763.02 | 456.57 |
| Grasse-Shanghai | 29 | 20 | 1030.85 | 893.98 | 605.99 | 823.41 | 531.02 |
| Grasse-Wuhan | 29 | 21 | 1021.22 | 883.47 | 560.09 | 867.04 | 489.76 |